\documentclass[aps,prd,showpacs,preprintnumbers,superscriptaddress,nofootinbib]{revtex4}
\usepackage[dvips]{graphicx}
\usepackage{bm,latexsym,amsmath,amssymb,amsfonts}
\newcommand{\stac}[2]{\stackrel{\scriptscriptstyle {#1}}{#2}}
\newcommand*{\py}{\partial_y}
\newcommand*{\cD}{{\cal D}}
\newcommand*{\cF}{{\cal F}}
\newcommand*{\cI}{{\cal I}}
\newcommand*{\cK}{{\cal K}}
\newcommand*{\e}{{\rm e}}
\begin{document}

\title{Low energy effective theory on a regularized brane in six-dimensional flux compactifications}

\author{Shunsuke~Fujii}
\email[Email: ]{fujii"at"th.phys.titech.ac.jp}
\affiliation{Department of Physics, Tokyo Institute of Technology, Tokyo 152-8551, Japan}
\author{Tsutomu~Kobayashi}
\email[Email: ]{tsutomu"at"gravity.phys.waseda.ac.jp}
\affiliation{Department of Physics, Waseda University, Okubo 3-4-1, Shinjuku, Tokyo 169-8555, Japan}
\author{Tetsuya~Shiromizu}
\email[Email: ]{shiromizu"at"phys.titech.ac.jp}
\affiliation{Department of Physics, Tokyo Institute of Technology, Tokyo 152-8551, Japan}

\begin{abstract}
Conical brane singularities in six-dimensional flux compactification models
can be resolved by introducing cylindrical codimension-one branes with regular caps
instead of 3-branes (a la Kaluza-Klein braneworlds with fluxes). 
In this paper, we consider such a regularized braneworld with axial symmetry
in six-dimensional Einstein-Maxwell theory. 
We derive a low energy effective theory on the regularized brane
by employing the gradient expansion approach,
and show that standard four-dimensional Einstein gravity is recovered at low energies.
Our effective equations extend to the nonlinear gravity regime,
implying that conventional cosmology can be reproduced in the regularized braneworld.
\end{abstract}

\preprint{WU-AP/272/07}
\maketitle
%%%%%%%%%%%%%%%%%%%%%%%%%%%%%%%%%%%%%%%%%%%%%%%%%%%%%%%%%%%%%%%%%%%%%%

\section{Introduction}

Recent developments in string theory provide us with a remarkably new picture of our 
universe, in which our universe is described by the motion of a 3-brane 
in a higher dimensional spacetime. In addition, it has been suggested that
extra dimensions warped due to self-gravity of branes may play a key role
in addressing the gauge hierarchy problem.
A simple toy model of warped braneworlds
was proposed by Randall and Sundrum~\cite{rs}
and its various consequences has been studied extensively~\cite{review}.
The idea of warped 
extra dimensions is plugged in warped flux compactifications, 
in which reliable inflation models based on string theory has been constructed~\cite{flux}
(see Ref.~\cite{flux-review} for a review). 

While the five-dimensional (5D) Randall-Sundrum-type 
models~\cite{rs} are
simple enough, offering us a feasible setup for
studying the behavior of gravity in braneworlds,
brane models in six or higher dimensions have a much richer structure
such as flux-stabilized compactifications.
In particular, codimension-two braneworlds with football-shaped extra dimensions
attracted much attention initially, due to their potential mechanism for
solving the cosmological constant problem~\cite{cc1, cc2, cc-rev}.
However, a higher codimension brane develops a conical or worse singularity
once taking into account its self-gravity, which
hampers a substantial analysis of higher dimensional braneworlds.
Thus, it is necessary to regularize higher codimension branes
in a well-controllable way to construct reasonable models
of higher dimensional braneworlds.
One of the regularization procedure is to replace
a higher codimension defect with a codimension-one brane,
whose extra dimensions are compactified in the traditional Kaluza-Klein manner. 
This was done in an specific model of 6D flux compactifications in~\cite{peloso1, ppz, peloso2}.
This type of models may be regarded as a variation of
Kaluza-Klein/hybrid braneworlds~\cite{hybrid, hybridns, takam, kssl, battye}.
Only pure tension, gravitational shockwaves, and black holes have so far been known to
be accommodated on codimension-two branes~\cite{bh}, but with this regularization scheme
one can put arbitrary energy-momentum tensor on the branes.
Performing perturbation analyses,
it has been shown that weak gravity sourced by arbitrary matter on the regularized brane
reproduces standard 4D one~\cite{peloso1, km}.

In this paper we further investigate the behavior of gravity
on the regularized brane in the 6D flux compactification models, generalizing the 
previous result to cover nonlinear gravity. 
We keep in mind the bulk-brane configuration of~\cite{peloso1, ppz},
but start with a more general metric ansatz.
Employing the gradient expansion approach,
we solve the 6D Einstein-Maxwell equations
to derive an effective theory on the brane.
The gradient expansion approach in braneworlds
is based on the $(4+1)$-decomposition of the bulk spacetime~\cite{sms}
and was originally developed in the context of the 5D Randall-Sundrum model~\cite{soda, sk}.
The gradient expansion has been widely used in various brane models (e.g.,~\cite{ge}).
Recently, along the direction of the regularized braneworlds,
the geometrical projection approach of~\cite{sms}
has been generalized to non-$Z_2$-symmetric models in an arbitrary number of
dimensions~\cite{ys}.

%In the present analysis, we assume an axisymmetric geometry
%and perform the $(5+1)$-decomposition of the 6D bulk,

The plan of the paper is as follows.
In the next section we perform the $(5+1)$-decomposition of 6D spacetime.
In Sec.~\ref{sec:GE} the evolution equations along one of the extra direction are solved by invoking
the gradient expansion approach up to first order.
We explain how conical singularities in the bulk are regularized by
introducing extended 4-branes in Sec.~\ref{sec:4BR}.
In Sec.~\ref{sec:EFF} we derive the effective theory on the regularized branes
using the result of Sec.~\ref{sec:GE}.
In Sec.~\ref{sec:CONCL} we draw our conclusions.

\section{Basic equations}\label{sec:BASIC}

The 6D action we consider is
\begin{eqnarray}
S =\int d^6x\sqrt{-g}\left[\frac{M^4}{2}
\left(R-\frac{1}{L_I^2}\right)-\frac{1}{4}F^2\right],
\end{eqnarray}
where $F^2=F_{MN}F^{MN}$ and $F_{MN}=\partial_MA_N-\partial_NA_M$
is the field strength of the $U(1)$ gauge field $A_M$.
$M$ is the fundamental scale of gravity and $1/2L_I^2$
is the cosmological constant.
Branes will be added in Sec.~\ref{sec:4BR}\footnote{
There we will glue the solutions with different cosmological constants together,
and so the subscript $I$ will eventually indicate the different regions of the spacetime.} and
for the time being we will concentrate on solving the 6D geometry.
We assume the axisymmetric metric ansatz in the form of
\begin{eqnarray}
g_{MN}dx^Mdx^N
=L_I^2\frac{dy^2}{f(y)}+\ell^2e^{2\psi(x, y)}f(y)d\theta^2+
2\ell b_{\mu}(x, y)d\theta dx^{\mu}+
\tilde g_{\mu\nu}(x, y)dx^{\mu}dx^{\nu},
\label{metric}
\end{eqnarray}
with
\begin{eqnarray}
\tilde g_{\mu\nu}=a^2(y)\tilde h_{\mu\nu}(x, y).
\end{eqnarray}
The capital Latin indices numerate the 6D coordinates, while
the Greek indices are restricted to the 4D coordinates.

We write the 6D Einstein equations in the form of the evolution equations
along the $y$-direction and the constraint equations.
The evolution equations are
\begin{eqnarray}
n^y\py K_{\hat\mu}^{\;\hat\nu}+\hat K K_{\hat\mu}^{\;\hat\nu}
={}^{5}\!R_{\hat\mu}^{\;\hat\nu}-\frac{1}{4L_I^2}\delta_{\hat\mu}^{\;\hat\nu}
-\frac{1}{M^4}\left(F_{\hat\mu M}F^{\hat\nu M}-\frac{1}{8}\delta_{\hat\mu}^{\;\hat\nu}F^2\right),
\end{eqnarray}
where $n^y=\sqrt{f}/L_I$, $K_{\hat\mu}^{\;\hat\nu}$ is the extrinsic curvature
of $y=$ constant hypersurfaces, and
$\hat K$ is its 5D trace.
${}^{5}\!R_{\hat\mu}^{\;\hat\nu}$ is the 5D Ricci tensor.
Here, $\hat\mu = \mu$ and $\theta$.
The Hamiltonian constraint is
\begin{eqnarray}
{}^{5}\!R+K_{\hat\mu}^{\;\hat\nu}K_{\hat\nu}^{\;\hat\mu}-\hat K^2
=\frac{1}{L_I^2}-\frac{2}{M^4}\left(F_{yM}F^{yM}-\frac{1}{4}F^2\right).
\end{eqnarray}
and
the momentum constraints are
\begin{eqnarray}
{}^{5}\! D_{\hat\nu}\left(K_{\hat\mu}^{\;\hat\nu}-\delta_{\hat\mu}^{\;\hat\nu}\hat K\right)
=\frac{1}{M^4}F_{\hat\mu M}F^{yM}n_y,
\end{eqnarray}
where ${}^{5}\! D_{\hat\mu}$ is the covariant derivative with respect to
the 5D metric and $n_y=L_I/\sqrt{f}$.

The Maxwell equations are given by
\begin{eqnarray}
\nabla_N F^{NM} =0,
\end{eqnarray}
where $\nabla_N$ is the covariant derivative with respect to the 6D metric. 

\section{Gradient expansion approach}\label{sec:GE}

We employ the gradient expansion approach to solve
the 6D bulk geometry~\cite{soda, sk}. Assuming that $\ell$ is not so different from $L_I$,
the small expansion parameter here is the ratio of the bulk curvature scale
to the 4D intrinsic curvature scale,
\begin{eqnarray*}
\varepsilon = \ell^2|R|.
\end{eqnarray*}
The metric and extrinsic curvature are expanded as
\begin{eqnarray*}
\tilde h_{\mu\nu}=h^{(0)}_{\mu\nu}+\varepsilon h^{(1)}_{\mu\nu}+\cdots,
\qquad
\psi=\psi^{(0)}+\varepsilon \psi^{(1)} +\cdots,
\\
K_{\mu}^{\;\nu}=\stac{(0)}{K_{\mu}^{\;\nu}}+
\;\varepsilon\!\!\stac{(1)}{K_{\mu}^{\;\nu}}+\cdots,
\qquad
K_{\theta}^{\;\theta}=\stac{(0)}{K_{\theta}^{\;\theta}}+
\;\varepsilon\!\!\stac{(1)}{K_{\theta}^{\;\theta}}+\cdots.
\end{eqnarray*}
We may assume naturally that
\begin{eqnarray}
b_{\mu}=\varepsilon^{1/2}b^{(1/2)}_{\mu}+\cdots,
\qquad K_{\theta}^{\;\nu}=\varepsilon^{1/2}
\stac{(1/2)}{K_{\theta}^{\;\nu}}+\cdots.\label{as1}
\end{eqnarray}
The $(y\theta)$ component of the field strength is expanded as
\begin{eqnarray*}
F_{y\theta}=\stac{(0)}{F_{y\theta}}+\varepsilon\!\!\stac{(1)}{F_{y\theta}}+\cdots.
\end{eqnarray*}
Since $\partial_{\mu}A_{\theta}\sim \varepsilon^{1/2} \py A_{\theta}$, we have
$F_{\mu\theta}=\varepsilon^{1/2}\!\!\stac{(1/2)}{F_{\mu\theta}}+\cdots$.
We also assume that
\begin{eqnarray}
F^{\mu y}=\varepsilon^{1/2}\!\!\stac{(1/2)}{F^{\mu y}}+\cdots,
\qquad
F_{\mu\nu}=\varepsilon \!\!\stac{(1)}{F_{\mu\nu}}+\cdots.\label{as2}
\end{eqnarray}
Therefore, the $F_{\mu\alpha}F^{\nu\alpha}$ part of the bulk energy-momentum tensor 
will be higher order in our treatment and it does not contribute to the 
low energy effective theory. The 5D Ricci tensor is given by 
\begin{eqnarray}
^{5}\! R_{\mu}^{\;\nu}&=& \varepsilon \frac{1}{a^2}\left(
 R_{\mu}^{\;\nu}[h^{(0)}]-\cD_{\mu}\cD^{\nu}\psi^{(0)}
-\cD_{\mu}\psi^{(0)}\cD^{\nu}\psi^{(0)}\right)+\cdots,
\\
^{5}\!R_{\theta}^{\;\theta}&=&-\varepsilon\frac{1}{a^2}\cD_{\lambda}\cD^{\lambda}\psi^{(0)}+\cdots,
\end{eqnarray}
and ${}^5\!R_{\theta}^{\;\mu}={\cal O}(\varepsilon^{3/2})$,\footnote{It turns out
in Appendix~\ref{app:o1/2} that we in fact have ${}^5\!R_{\theta}^{\;\mu}={\cal O}(\varepsilon^{5/2})$.} where
$R_{\mu}^{\;\nu}[h^{(0)}]$ stands for the 4D Ricci tensor of the zeroth order metric $h^{(0)}_{\mu\nu}$
and $\cD_{\mu}$ is the covariant derivative with respect to $h^{(0)}_{\mu\nu}$.

\subsection{Zeroth order equations}

At zeroth order in the gradient expansion, the evolution equations and the Hamiltonian constraint are
\begin{eqnarray}
\frac{\sqrt{f}}{L_I}\py \stac{(0)}{K_{\mu}^{\;\nu}}
+\Bigl(
\stac{(0)}{K_{\lambda}^{\;\lambda}}+\stac{(0)}{K_{\theta}^{\;\theta}}
\Bigr)\stac{(0)}{K_{\mu}^{\;\nu}}&=&
-\frac{1}{4L_I^2}\left[1-\frac{1}{\ell^2M^4}\Bigl(
e^{-\psi^{(0)}}
\stac{(0)}{F_{y\theta}}\Bigr)^2\right]\delta_{\mu}^{\;\nu},
\\
\frac{\sqrt{f}}{L_I}\py \stac{(0)}{K_{\theta}^{\;\theta}}
+\Bigl(
\stac{(0)}{K_{\lambda}^{\;\lambda}}+\stac{(0)}{K_{\theta}^{\;\theta}}
\Bigr)\stac{(0)}{K_{\theta}^{\;\theta}}&=&
-\frac{1}{4L_I^2}\left[1+\frac{3}{\ell^2M^4}\Bigl(
e^{-\psi^{(0)}}
\stac{(0)}{F_{y\theta}}\Bigr)^2\right],
\\
\stac{(0)}{K_{\mu}^{\;\nu}}\stac{(0)}{K_{\nu}^{\;\mu}}
-\Bigl(\stac{(0)}{K_{\lambda}^{\;\lambda}}\Bigr)^2
-2\stac{(0)}{K_{\lambda}^{\;\lambda}}\stac{(0)}{K_{\theta}^{\;\theta}}
&=&
\frac{1}{L_I^2}\left[1-\frac{1}{\ell^2M^4}\Bigl(
e^{-\psi^{(0)}}
\stac{(0)}{F_{y\theta}}\Bigr)^2\right].
\end{eqnarray}
The Maxwell equations at zeroth order is
\begin{eqnarray}
\py\Bigl(
\sqrt{-h^{(0)}}a^4e^{-\psi^{(0)}}\!\!\stac{(0)}{F_{y\theta}}
\Bigr)=0.
\end{eqnarray}

The above equations are solved by
\begin{eqnarray}
a(y) &=& y,\label{warpf}\\
f(y)&=&-\frac{y^2}{20}+\frac{\mu}{y^3}-\frac{Q^2}{12y^6},
\label{backgroundsol}
\end{eqnarray}
and
\begin{eqnarray}
\stac{(0)}{F_{y\theta}}=\ell M^2\frac{Q}{y^4}e^{\psi^{(0)}},
\quad
h^{(0)}_{\mu\nu}(x, y)=h_{\mu\nu}(x),
\quad
\psi^{(0)}(x, y) = \psi^{(0)}(x),
\label{zerothsol}
\end{eqnarray}
with $\mu$ and $Q$ being constants that characterize the geometry of
the internal space.

We assume that the metric function $f(y)$ has two positive roots, $y_N$ and $y_S(<y_N)$. 
In other words, we choose the parameters $\mu$ and $Q$ so that $f(y)$ has two 
positive roots. 
We impose that $A_{\theta}$ vanishes at these poles, yielding
(in the northern half)
\begin{eqnarray}
A^{(0)}_{\theta}=\frac{\ell M^2Q}{3}\left(\frac{1}{y_N^3}-\frac{1}{y^3}\right)e^{\psi^{(0)}}.
\end{eqnarray}
From this we obtain $\stac{(1/2)}{F_{\mu\theta}}=\partial_{\mu}\psi^{(0)}\cdot A_{\theta}^{(0)}$.
Substituting this into the $\mu$ component of
the momentum constraints, we have $\partial_{\mu}\psi^{(0)}=0$.
Therefore, $\psi^{(0)}$ must be constant and
without loss of generality we may set $\psi^{(0)}=0$.
Furthermore, this implies $\stac{(1/2)}{F_{\mu\theta}}=0$ and hence
$F_{\mu\theta}= {\cal O}(\varepsilon^{3/2})+\cdots$.% as seen in Appendix A.

The above bulk geometry is our ``background,'' which is essentially
the double Wick rotated Reissner-Nordstr\"{o}m solution~\cite{Mukohyama, hybrid}.

\subsection{First order equations}

Going to first order in the gradient expansion
we will obtain the gravitational field equations that govern the
behavior of the 4D metric $h^{(0)}_{\mu\nu}$.
The first order equations may contain terms like $\sim \stac{(1/2)}{F_{\mu y}}\stac{(1/2)}{F^{\nu y}}$
and $\sim \stac{(1/2)}{K_{\theta}^{\;\nu}}\stac{(1/2)}{K_{\nu}^{\;\theta}}$.
However, in Appendix~\ref{app:o1/2}
we analyze the ${\cal O}(\varepsilon^{1/2})$ equations and show that
the gradient expansion of $F_{\mu y}$ and $K_{\theta}^{\;\nu}$ in fact begins with
${\cal O}(\varepsilon^{3/2})$.
Therefore, in the following we will drop the contributions from such terms.

At first order, the $(\mu\nu)$ component of the evolution equations is given by
\begin{eqnarray}
\frac{\sqrt{f}}{L_I}\left[
\py \!\stac{(1)}{K_{\mu}^{\;\nu}}+
\left(\frac{4}{y}+\frac{\py f}{2f}\right)\!\stac{(1)}{K_{\mu}^{\;\nu}}
+\frac{1}{y}\Bigl(
\stac{(1)}{K_{\lambda}^{\;\lambda}}+\stac{(1)}{K_{\theta}^{\;\theta}}
\Bigr)\delta_{\mu}^{\;\nu}
\right]
=\frac{1}{y^2}R_{\mu}^{\;\nu}
+\frac{1}{4}\cF\delta_{\mu}^{\;\nu},
\label{ev1st}
\end{eqnarray}
where we defined the useful combination
\begin{eqnarray}
\cF:=\frac{1}{M^4}\Bigl(\stac{(0)}{F_{y\theta}}\stac{(1)}{F^{y\theta}}
+\stac{(1)}{F_{y\theta}}\stac{(0)}{F^{y\theta}}\Bigr).
\end{eqnarray}
The 4D Ricci tensor $R_{\mu}^{\;\nu}$ does not depend on $y$
because it is computed from $h^{(0)}_{\mu\nu}$ which is a function of only $x^{\mu}$.
The traceless part of Eq.~(\ref{ev1st}) is found to be
\begin{eqnarray}
\py\left(y^4\sqrt{f}\;\mathbb{K}_{\mu}^{\;\nu}\right)
= y^2L_I\mathbb{R}_{\mu}^{\;\nu},
\label{tl}
\end{eqnarray}
where
we defined the traceless part of the relevant tensors as
$\mathbb{K}_{\mu}^{\;\nu}:=\stac{(1)}{K_{\mu}^{\;\nu}}-(1/4)\delta_{\mu}^{\;\nu}\stac{(1)}{K_{\lambda}^{\;\lambda}}$
and
$\mathbb{R}_{\mu}^{\;\nu}:=R_{\mu}^{\;\nu}-(1/4)\delta_{\mu}^{\;\nu}R$.
Eq.~(\ref{tl}) can be integrated to give
\begin{eqnarray}
\mathbb{K}_{\mu}^{\;\nu}=\frac{1}{3y\sqrt{f}}L_I\mathbb{R}_{\mu}^{\;\nu}
+\frac{1}{y^4\sqrt{f}}\mathbb{C}_{\mu}^{\;\nu}(x),
\end{eqnarray}
where the traceless tensor $\mathbb{C}_{\mu}^{\;\nu}(x)$ is the integration ``constant''
to be fixed by the boundary conditions.

The trace part of the evolution equations is
\begin{eqnarray}
\frac{\sqrt{f}}{L_I}\left[
\py \!\stac{(1)}{K_{\lambda}^{\;\lambda}}+\left(\frac{8}{y}+\frac{\py f}{2f}\right)
\!\stac{(1)}{K_{\lambda}^{\;\lambda}}+\frac{4}{y}\stac{(1)}{K_{\theta}^{\;\theta}}
\right]=\frac{1}{y^2}R+\cF,
\label{tp1st}
\end{eqnarray}
and the $(\theta\theta)$ component of the evolution equations is
\begin{eqnarray}
\frac{\sqrt{f}}{L_I}\left[
\py \!\stac{(1)}{K_{\theta}^{\;\theta}}+
\left(\frac{4}{y}+\frac{\py f}{f}\right)\!\stac{(1)}{K_{\theta}^{\;\theta}}
+\frac{\py f}{2f}\stac{(1)}{K_{\lambda}^{\;\lambda}}
\right]
=-\frac{3}{4}\cF.
\label{thth1st}
\end{eqnarray}
The Hamiltonian constraint at first order reduces to
\begin{eqnarray}
\frac{1}{y^2}R+\cF
=2\frac{\sqrt{f}}{L_I}\left[
\left(\frac{3}{y}+\frac{\py f}{2f}\right)\stac{(1)}{K_{\lambda}^{\;\lambda}}+
\frac{4}{y}\stac{(1)}{K_{\theta}^{\;\theta}}\right].
\label{ham1st}
\end{eqnarray}
It is possible to solve
the set of equations~(\ref{tp1st})--(\ref{ham1st}) and
find the bulk profile of the extrinsic curvature and $\cF$ (see Appendix~\ref{app:solve}).
Then, with the help of the relation
\begin{eqnarray}
\stac{(1)}{F_{y\theta}}=\ell^2L_I^2\Bigl(2\psi^{(1)}\!\stac{(0)}{F^{y\theta}}+
\stac{(1)}{F^{y\theta}}\Bigr),\label{FFF}
\end{eqnarray}
one finds the bulk profile of the gauge field.
Equivalently, one can use
the first order Maxwell equations,
\begin{eqnarray}
\py\Bigl(y^4\stac{(1)}{F^{y\theta}}\Bigr)+\frac{L_I}{\sqrt{f}} \,y^4\!\! \stac{(0)}{F^{y\theta}}
\Bigl(
\stac{(1)}{K_{\lambda}^{\;\lambda}}+\stac{(1)}{K_{\theta}^{\;\theta}}\Bigr)=0,
\label{max1st}
\end{eqnarray}
to solve for the gauge field.
As will be seen, however, we do not need
the explicit form of the solution for the extrinsic curvature and gauge field
in order to derive an effective theory on the brane;
it suffices to solve the bulk evolution of the combination of first order variables
\begin{eqnarray}
\cK:=\frac{3}{4}\stac{(1)}{K_{\lambda}^{\;\lambda}}+\stac{(1)}{K_{\theta}^{\;\theta}}
+\frac{1}{M^4}\frac{L_I}{\sqrt{f}}\stac{(0)}{F^{y\theta}}\!\!A_{\theta}^{(1)}
+\frac{\sqrt{f}}{L_I}\left(\frac{\py f}{2f}-\frac{1}{y}\right)\psi^{(1)}.
\end{eqnarray}
The evolution equation for $\cK$ can be obtained as follows.
Using the relation~(\ref{FFF}),
it is straightforward to show
\begin{eqnarray}
\py\left(y^4\sqrt{f}\cK\right)=
\py\left[
y^4\sqrt{f}\left(\frac{3}{4}\stac{(1)}{K_{\lambda}^{\;\lambda}}+\stac{(1)}{K_{\theta}^{\;\theta}}\right)
\right]
+y^4\sqrt{f}\left(\frac{\py f}{2f}-\frac{1}{y}\right)\stac{(1)}{K_{\theta}^{\;\theta}}
+\frac{1}{2}y^4L_I\cF.
\end{eqnarray}
Then, using the evolution equations~(\ref{tp1st}),~(\ref{thth1st}), and the Hamiltonian
constraint~(\ref{ham1st}), we arrive at
\begin{eqnarray}
\py\left(y^4\sqrt{f}\cK\right)=\frac{y^2}{4}L_I R.
\label{treq}
\end{eqnarray}
The structure of Eq.~(\ref{treq}) is identical to the traceless equations~(\ref{tl}),
and the general solution is given by
\begin{eqnarray}
\cK=\frac{1}{12y\sqrt{f}}L_IR+\frac{1}{y^4\sqrt{f}}\chi(x),
\end{eqnarray}
where $\chi(x)$ is an integration ``constant'' to be determined by the boundary conditions.

The $\mu$ component of the momentum constraints at first order is given by
\begin{eqnarray*}
\cD_{\nu}\mathbb{K}_{\mu}^{\;\nu}-\cD_{\mu}\left(\frac{3}{4}
\stac{(1)}{K_{\lambda}^{\;\lambda}}+\stac{(1)}{K_{\theta}^{\;\theta}}
\right)+\frac{\sqrt{f}}{L_I}\left(\frac{1}{y}-\frac{\py f}{2f}\right)\cD_{\mu}\!\!\stac{(1)}{\psi}
=\frac{1}{M^4}\cD_{\mu}\!\!\stac{(1)}{A_{\theta}}\;\stac{(0)}{F^{y\theta}}\frac{L_I}{\sqrt{f}},
\end{eqnarray*}
which
can be simplified using $\cK$ to
\begin{eqnarray}
\cD_{\nu}\mathbb{K}_{\mu}^{\;\nu}-\cD_{\mu}\cK=0.
\label{mom1st}
\end{eqnarray}
Eq.~(\ref{mom1st}) and the Bianchi identity,
$\cD_{\nu}\left[R_{\mu}^{\;\nu}-(1/2)\delta_{\mu}^{\;\nu}R\right]=0$,
imply the constraint for the integration constants:
\begin{eqnarray}
\cD_{\nu}\mathbb{C}_{\mu}^{\;\nu}-\cD_{\mu}\chi=0.\label{c-chi}
\end{eqnarray}

\section{Replacing conical branes with regularized 4-branes}\label{sec:4BR}

The metric function $f(y)$ vanishes at $y_N$ and $y_S$.
These points develop conical singularities in general and they are regarded as source 3-branes.
To resolve the singularities,
we replace each of the conical branes with a cylindrical codimension-one brane
and fill in the interior with a regular cap~\cite{peloso1, ppz}.
The geometries of the two caps and central bulk region are described by
the 6D solutions obtained in the previous section with different cosmological constants
($L_+$ ($L_-$) for the north (south) cap and $L_0$ for the central bulk).
Near the pole $y=y_p$ ($p=N, S$), we have $f(y)\simeq \py f|_{y=y_p}(y-y_p)$.
In order for the cap to close regularly at $y_p$, we impose
\begin{eqnarray}
\left.\frac{\ell}{L_I}\frac{|\py f|}{2}e^{\psi}\right|_{y=y_p}=2\pi.
\end{eqnarray}
Clearly, it is required that $\psi(x, y_p)=$ constant.
Without loss of generality we can set
this constant contribution to be zero.
In particular, we have $\psi^{(1)}\lesssim (y-y_p)$ near $y=y_p$.

We consider the following action for each 4-brane~\cite{peloso1}:
\begin{eqnarray}
S^{\pm}=-\int d^5x\sqrt{-q}\left[\lambda_{\pm}+\frac{v_{\pm}^2}{2}q^{\hat\mu\hat\nu}
(\partial_{\hat\mu}\Sigma_{\pm}-\e A_{\hat\mu})
(\partial_{\hat\nu}\Sigma_{\pm}-\e A_{\hat\nu})\right]+S_m^{\pm},
\end{eqnarray}
where $\lambda_{\pm}$ is the brane tension,
$v_{\pm}$ is the vacuum expectation value of the brane Higgs field
and $\Sigma_{\pm}$ is its Goldstone mode. $S^{\pm}_m$ represents the matter action on the brane.
The brane action necessarily couples to $A_{\hat\mu}$ in order to account
for the jump of the Maxwell field at the brane. This implies that the 
brane location is fixed[see Eq.~(\ref{mj0}) below]. 
The total brane energy-momentum tensor derived from the above action is
\begin{eqnarray}
T_{\hat\mu(tot)}^{\pm \hat\nu}=-\lambda_\pm \delta_{\hat\mu}^{\;\hat\nu}+v^2_\pm
\left[
\left(\partial_{\hat\mu}\Sigma_\pm-\e A_{\hat\mu}\right)
\left(\partial^{\hat\nu}\Sigma_\pm-\e A^{\hat\nu}\right)
-\frac{1}{2}\left(\partial_{\hat\lambda}\Sigma_\pm-\e A_{\hat\lambda}\right)
\left(\partial^{\hat\lambda}\Sigma_\pm-\e A^{\hat\lambda}\right)
\delta_{\hat\mu}^{\;\hat\nu}
\right]+T_{\hat\mu}^{\pm \hat\nu},
\end{eqnarray}
where $T_{\hat\mu}^{\pm \hat\nu}$ is the energy-momentum tensor of the matter fields 
on the branes.
We assume that $T_{\mu}^{\pm \theta}=T_{\theta}^{\pm  \nu}=0$.

Each of the caps and the central bulk spacetime are glued together
so as to satisfy the Israel conditions~\cite{Israel} and the jump conditions for the Maxwell field.
From now on we will suppress the index $\pm$ if it is not necessary.
We start with assuming
that the brane location
is given by a $x$-dependent function $y=\varphi (x)$.
The brane induced metric is given by
$q_{\hat\mu\hat\nu}dx^{\hat\mu}dx^{\hat\nu}=\ell^2f(\varphi(x))d\theta^2+\varphi^2(x)h^{(0)}_{\mu\nu}dx^{\mu}dx^{\nu}+{\cal O}(\varepsilon)$.
The equation of motion for each scalar field $\Sigma_{\pm}$
at zeroth order reduces to $\partial_{\theta}^2\Sigma^{(0)}_{\pm}=0$,
and hence
\begin{eqnarray}
\Sigma^{(0)}_{\pm}=n_{\pm}\theta+\sigma_{\pm}^{(0)}(x).\label{bH0}
\end{eqnarray}
Here, $n_{\pm}$ must be integer
because $\Sigma_{\pm}$ is the phase of the Higgs field
and so $e^{i\Sigma(\theta+2\pi, x)}=e^{i\Sigma(\theta, x)}$.
This property should hold at any order in the gradient expansion.
Therefore, the solution for $\Sigma$ including higher order corrections should be of the form
$\Sigma=n\theta+\sigma^{(0)}(x)+\varepsilon\sigma^{(1)}(x)+\cdots$.

We shall show now that the brane location is in fact independent of $x$. 
To do so,
let us consider the $\theta$ component of the Maxwell jump conditions at zeroth order:
\begin{eqnarray}
\Bigl[\Bigl[ \bar n^M\!\!\stac{(0)}{F_{M\theta}}\Bigr]\Bigr]=
-\e v^2\left(\partial_{\theta}\Sigma^{(0)}-\e A_{\theta}^{(0)}\right),\label{mj0}
\end{eqnarray}
where
$[[A ]]:=A|_{y=\varphi+\epsilon}-A|_{y=\varphi-\epsilon}$ and
$\bar n^M$ is the unit normal to the brane.
At this order, we have $\bar n^M \simeq (n^y, 0)$.
The right hand side is independent of $x$ and hence
the brane location must be $y=$ constant $(=: y_{\pm})$.
The Israel conditions are given by
\begin{eqnarray}
\left[\left[
\frac{\sqrt{f}}{L_I}\left(\frac{3}{y}+\frac{\py f}{2f}\right)
\right]\right]&=&-\frac{\lambda}{M^4}-\frac{v^2}{2M^4}\Big(\partial_{\theta}\Sigma^{(0)}
-\e A^{(0)}_{\theta}\Big)
\Big(\partial^{\theta}\Sigma^{(0)}-\e A^{\theta(0)}\Big)
=T_{\mu(tot)}^{\;\nu(0)},\label{is01}
\\
\left[\left[
\frac{\sqrt{f}}{L_I}\frac{4}{y}
\right]\right]&=&-\frac{\lambda}{M^4}+\frac{v^2}{2M^4}\Big(\partial_{\theta}\Sigma^{(0)}
-\e A^{(0)}_{\theta}\Big)
\Big(\partial^{\theta}\Sigma^{(0)}-\e A^{\theta(0)}\Big)
=T_{\theta(tot)}^{\;\theta (0)}.\label{is02}
\end{eqnarray}
The above conditions~(\ref{mj0})--(\ref{is02})
determine the brane location $y_{\pm}$ and the parameter of the solution $n_{\pm}$.
For the current purpose
we need no further detail;
see Sec.~II of Ref.~\cite{km} for more explanation of the configuration of the branes.

%Now it is clear that the St\"{u}ckelberg term
%tends to cancel the pressure in the $\theta$ direction.

%Since there are six conditions (three for each brane),
%the parameters in the brane action, $\{\lambda_{\pm}, v_{\pm}, \e\}$,
%cannot be independent of each other.

%There are apparently seven
%parameters for the zeroth order configuraiton: $\{y_{\pm}, n_{\pm}, \ell, \mu, Q\}$.
%We can show, however, that using $\alpha:=y_S/y_N$ and $\bar \ell:=y_N \ell$ instead of 
%$\ell$, $\mu$, and $Q$ is enough to characterize the configuration.
%Hence, the relevant parameters of the solution are $\{y_{\pm}, n_{\pm}, \bar\ell, \alpha\}$.

%Given the parameters in the action, $\{v_{\pm}, \e, \lambda_{\pm}, L_I\}$,
%and those characterizing the bulk geometry, $\{\mu, Q, \ell\}$,
%Eqs.~(\ref{is01}) and~(\ref{is02}) are solved to give $y_{\pm}$ and $n_{\pm}$
%as functions of these parameters.

\section{Effective theory on a regularized brane}\label{sec:EFF}

We go on to
specifying the first order boundary conditions at the poles and branes.
As to the regularity conditions at the poles,
it is required that
\begin{eqnarray}
\mathbb{K}_{\mu}^{\;\nu},\;\;
\stac{(1)}{K_{\lambda}^{\;\lambda}},\;\;
\stac{(1)}{K_{\theta}^{\;\theta}} \;\;\lesssim  |y-y_p|^{1/2}.
\end{eqnarray}
With this, the evolution equations for the extrinsic
curvature~(\ref{tl}), (\ref{tp1st}), and (\ref{thth1st})
are regular at the poles.
We also require that $|\stac{(1)}{F_{y\theta}}|<\infty$ at $y=y_p$.
This implies that $A^{(1)}_{\theta}\lesssim  (y-y_p)$ near the pole.
Noting that $\psi^{(1)}\lesssim(y-y_p)$,
we have $\cK\lesssim  |y-y_p|^{1/2}$ near $y= y_p$.

The $(\mu\nu)$ component of the Israel conditions at first order are given by
\begin{eqnarray}
 \Bigl[\Bigl[\stac{(1)}{K_{ \mu}^{\; \nu}}-\delta_{ \mu}^{\; \nu}
\Bigl(
\stac{(1)}{K_{\lambda}^{\;\lambda}}+\stac{(1)}{K_{\theta}^{\;\theta}}
\Bigr)\Bigr]\Bigr]
=-\frac{T_{ \mu}^{\; \nu}}{M^4}
-\frac{ v^2}{M^4}
\frac{1}{\ell^2 f}\left[
\Bigl(\partial_{\theta}\Sigma^{(0)}-\e  A_{\theta}^{(0)}\Bigr)\e A^{(1)}_{\theta}
+\Bigl(\partial_{\theta}\Sigma^{(0)}-\e A_{\theta}^{(0)}\Bigr)^2\psi^{(1)}
\right]\delta_{ \mu}^{\; \nu}.\label{ic1st}
\end{eqnarray}
In deriving this we used
$\left(\partial^{\nu}\Sigma-\e A^{\nu}\right)^{(1/2)}=0$ which is shown in Appendix~\ref{app:o1/2}.
The traceless part of Eq.~(\ref{ic1st}) reads
\begin{eqnarray}
\left[\left[\mathbb{K}_{\mu}^{\;\nu}\right]\right]=-\frac{1}{M^4}\mathbb{T}_{\mu}^{\;\nu},
\label{jctl}
\end{eqnarray}
where $\mathbb{T}_{\mu}^{\;\nu}$ is the traceless part of the energy-momentum tensor.
Noting that\footnote{
The gauge field $A_{\theta}^{(1)}$ is continuous across the brane
in order for the brane action to be well-defined.
The continuity of the induced metric imposes that
$\psi^{(1)}$ is continuous across the brane.}
\begin{eqnarray*}
[[\cK]]=\frac{1}{4}\Bigl[\Bigl[3\stac{(1)}{K_{\lambda}^{\;\lambda}}+4
\stac{(1)}{K_{\theta}^{\;\theta}}\Bigr]\Bigr]
+\frac{1}{M^4}\frac{1}{\ell^2 f}\Bigl[\Bigl[
n^y\stac{(0)}{F_{y\theta}}
\Bigr]\Bigr] A_{\theta}^{(1)}
+\left[\left[
\frac{\sqrt{f}}{L_I}\left(\frac{\py f}{2f}-\frac{1}{y}\right)
\right]\right]\psi^{(1)},
\end{eqnarray*}
and using the zeroth order Israel conditions and jump conditions
for the Maxwell field,
we find that the trace part of Eq.~(\ref{ic1st}) yields
\begin{eqnarray}
\left[\left[\cK\right]\right]=\frac{1}{4M^4}T_{\lambda}^{\;\lambda}.\label{jct}
\end{eqnarray}
The junction conditions~(\ref{jctl}) and~(\ref{jct}) together with
the momentum constraint~(\ref{mom1st}) imply the local conservation law for the 
energy-momentum tensor of brane localized matter:
\begin{eqnarray}
\cD_{\nu}T_{\mu}^{\;\nu}=0.
\end{eqnarray}

Now we are to fix the integration constants.
From the regularity condition at the north pole one can determine
the integration constant in the north cap and we have
\begin{eqnarray}
\mathbb{K}_{\mu}^{\;\nu}=\frac{y^3-y^3_N}{3y^4\sqrt{f}}L_+\mathbb{R}_{\mu}^{\;\nu}
\quad(y_+<y\leq y_N).
\end{eqnarray}
Then, using the Israel conditions at the brane we obtain
\begin{eqnarray}
\mathbb{K}_{\mu}^{\;\nu}=\frac{1}{3y\sqrt{f}}L_0\mathbb{R}_{\mu}^{\;\nu}+\frac{1}{y^4\sqrt{f}}
\mathbb{C}_{\mu}^{\;\nu} \quad
(y_-<y<y_+),
\end{eqnarray}
where
\begin{eqnarray}
\mathbb{C}_{\mu}^{\;\nu}=-\frac{(y_N^2-y_+^3)L_++y_+^3L_0}{3}\mathbb{R}_{\mu}^{\;\nu}
+y_+^4\sqrt{f_+}\frac{\mathbb{T}_{\mu}^{+\nu}}{M^4}
\end{eqnarray}
and $f_+:=f(y_+)$. 
The regularity at the south pole determines the extrinsic curvature in the south cap as
\begin{eqnarray}
\mathbb{K}_{\mu}^{\;\nu}=\frac{y^3-y^3_S}{3y^4\sqrt{f}}L_-\mathbb{R}_{\mu}^{\;\nu}
\quad(y_S\leq y< y_-).
\end{eqnarray}
The Israel conditions at $y=y_-$ imply
\begin{eqnarray}
\frac{(y_-^3-y_S^3)L_--y_-^3L_0}{3}\mathbb{R}_{\mu}^{\;\nu}-\mathbb{C}_{\mu}^{\;\nu}
=y_-^4\sqrt{f_-}\frac{\mathbb{T}_{\mu}^{-\nu}}{M^4},
\end{eqnarray}
which is rearranged to give
\begin{eqnarray}
\frac{\ell_*^2}{\ell}\mathbb{R}_{\mu}^{\;\nu}=\frac{1}{M^4}\left(
y_+^4\sqrt{f_+}\mathbb{T}_{\mu}^{+\nu}
+y_-^4\sqrt{f_-}\mathbb{T}_{\mu}^{-\nu}
\right),\label{tr:einstein}
\end{eqnarray}
where
\begin{eqnarray}
\ell_*^2:=\ell\int^{y_N}_{y_S}\!\!L_Iy^2dy
\end{eqnarray}
and $f_-:=f(y_-)$. Obviously, for the trace part we have the same relation as~(\ref{tr:einstein})
with the substitution $\mathbb{R}_{\mu}^{\;\nu}\to R/4$
and $\mathbb{T}_{\mu}^{\pm\nu}\to -T^{\pm\lambda}_{\lambda}/4$
(and thus the constraint for the integration constants~(\ref{c-chi}) is trivially satisfied).
Taking into account that
the induced metric on the north brane is given by $y_+^2h_{\mu\nu}$,
we define the brane Ricci tensor as ${\cal R}_{\mu}^{\;\nu}:=R_{\mu}^{\;\nu}/y_+^2$.
Combining the traceless and trace equations, we finally arrive at
\begin{eqnarray}
{\cal R}_{\mu}^{\;\nu}-\frac{1}{2}\delta_{\mu}^{\;\nu}{\cal R}=
\kappa^2_+ \overline{T}_{\mu}^{+\nu}+\frac{y_-^2}{y_+^2}\kappa^2_-\overline{T}_{\mu}^{-\nu},
\label{einstein}
\end{eqnarray}
where we defined the 4D gravitational coupling at each brane as
\begin{eqnarray}
\kappa^2_{\pm}:=\frac{y_{\pm}^2}{2\pi\ell_*^2 M^4},
\end{eqnarray}
and the energy-momentum tensor integrated along the $\theta$-direction as
$\overline{T}_{\mu}^{\pm\nu}:=2\pi\ell \sqrt{f_\pm}T_{\mu}^{\pm\nu}$.
Eq.~(\ref{einstein}) shows that gravity at low energies is described by
general relativity (when matter on the south brane can be neglected.)
This generalizes the perturbative analysis of~\cite{peloso1, km}
to the nonlinear regime.
The effect of the extra scalar mode that appears in~\cite{peloso1, km}
is higher order in the gradient expansion and hence is not observed
in the present leading order analysis.

\section{Conclusions}\label{sec:CONCL}

In this paper we have considered six-dimensional Einstein-Maxwell model of
warped braneworlds, where the extra dimensions are stabilized by a flux.
Conical singularities in such models can be regularized by replacing the 3-branes with
codimension-one branes.
Using the gradient expansion approach,
we have derived a low energy effective theory on the regularized brane,
and have shown that standard 4D Einstein gravity is reproduced on the brane.
Though the present analysis is restricted up to
first order, one can go on to the higher order expansions to include
the effect of Kaluza-Klein modes.
The scalar zero mode corresponding to the fluctuation of the internal space volume
would also appear at second order.
Computing higher order corrections would be interesting
and is one of the remaining issues.

Our result has been derived without relying on
linear perturbations in the Minkowski brane background.
The effective Einstein equations~(\ref{einstein}) are valid even
in the nonlinear regime,
and hence can be applied to the cosmological dynamics at low energies (i.e., to late-time cosmology
in which the Hubble horizon is much larger than the compactification scale).
Since the effective theory coincides with 4D general relativity,
the present model can reproduce conventional cosmology on the brane.
Recently,
cosmology on a moving brane in a {\em given (static)} background
has been investigated in the context of regularized braneworlds with flux compactifications~\cite{ppz_cos, ml_cos}.
Their results are not compatible
with the realistic cosmological evolution,
suggesting that the static bulk assumption is an oversimplification.
On the other hand, we have started with a more general metric ansatz~(\ref{metric})
whose internal space component is nontrivially dependent on the external spacetime coordinates.
Thus, constructing explicit cosmological solutions in the 6D flux compactification model
is not so simple as in the 5D Randall-Sundrum-type braneworlds
(see, however, \cite{Cline:2003ak, 6d_cos}).

6D braneworlds are often discussed
not only in a simplified Einstein-Maxwell setup but also
in the context of gauged chiral supergravity~\cite{sugra, sugra2}.
Supergravity models generally have conical brane singularities too,
and the regularization procedure has been proposed in~\cite{ppz, uvcap}.
The present gradient expansion approach is useful for the supergravity braneworlds as well.

%**************************************************

%--- Acknowledgements ---%--- Acknowledgements ---%--- Acknowledgements ---%
\acknowledgments
SF and TK are supported by the JSPS under Contract Nos.~19-2570 and 19-4199. 
TS is supported by Grant-Aid for Scientific Research from Ministry of 
Education, Science, Sports and Culture of Japan (Nos.~17740136, 17340075, and 19GS0219), 
the Japan-U.K., Japan-France and Japan-India Research Cooperative Programs.

%--- Acknowledgements ---%--- Acknowledgements ---%--- Acknowledgements ---%

\appendix

\section{${\cal O}(\varepsilon^{1/2})$ quantities}\label{app:o1/2}

Starting with the assumptions~(\ref{as1}) and~(\ref{as2}), we shall show in this appendix that
the leading order terms in the gradient expansion of $K_{\theta}^{\;\nu}$ and $F^{\mu y}$ are
in fact ${\cal O}(\varepsilon^{3/2})$.

The $\mu$ component of the ${\cal O}(\varepsilon^{1/2})$ Maxwell equations reads
\begin{eqnarray}
\py\Bigl(y^4 \stac{(1/2)}{F^{y\mu}}\Bigr) = 0,
\end{eqnarray}
and thus we have
\begin{eqnarray}
\stac{(1/2)}{F^{\mu y}} = M^2\frac{C_1^{\mu}(x)}{y^4}.
\end{eqnarray}
The ${\cal O}(\varepsilon^{1/2})$ evolution equation reduces to
\begin{eqnarray}
\frac{1}{L_I}\frac{1}{y^4}\py\Bigl(y^4\sqrt{f}\!\stac{(1/2)}{K_{\theta}^{\;\nu}}\Bigr)
&=&-\frac{1}{M^4}\!\stac{(0)}{F_{\theta y}}\stac{(1/2)}{F^{\nu y}}
\nonumber\\
&=& \ell Q\frac{C_1^{\nu}(x)}{y^8},
\end{eqnarray}
which can be integrated to give
\begin{eqnarray}
\stac{(1/2)}{K_{\theta}^{\;\nu}} = \frac{1}{y^4\sqrt{f}}\left[
-\frac{L_I\ell Q}{3}\frac{C_1^{\nu}(x)}{y^3}+C_2^{\nu}(x)
\right].
\end{eqnarray}

The ${\cal O}(\varepsilon)$ evolution equations
contain terms like
$\stac{(1/2)}{F_{\mu y}}\stac{(1/2)}{F^{\nu y}}\propto h_{\mu\lambda}C_1^{\lambda}C_1^{\nu}/[y^6 f(y)]$.
Thus,
in the cap regions it is required that $C_1^{\nu}=0$ because
otherwise this term would show a singular behavior at the poles.
Similarly, to ensure the regular behavior of $\stac{(1/2)}{K_{\theta}^{\;\nu}}$ at the poles
we impose $C_2^{\nu}=0$ in the cap regions.
To fix the integration constants in the central bulk region,
we invoke the junction conditions at the branes.
The jump conditions of the Maxwell field and Israel conditions imply, respectively,
\begin{eqnarray}
\Bigl[\Bigl[n_y\stac{(1/2)}{F^{y\mu}}\Bigr]\Bigr] &=& -\e v^2
\left(\partial^{\mu}\Sigma-\e A^{\mu}\right)^{(1/2)},
\\
\Bigl[\Bigl[\stac{(1/2)}{K_{\theta}^{\;\mu}}\Bigr]\Bigr] &=& -\frac{ v^2}{M^4}
\left(\partial_{\theta} \Sigma^{(0)}-\e A_{\theta}^{(0)}\right)
\left(\partial^{\mu}\Sigma-\e A^{\mu}\right)^{(1/2)}.
\end{eqnarray}
Combining these two equations and noting that
$\stac{(1/2)}{K_{\theta}^{\;\mu}}=0=\stac{(1/2)}{F^{y\mu}}$ in each cap, we obtain
two linear algebraic equations for $C_1^{\nu}$ and $C_2^{\nu}$:
\begin{eqnarray}
\left.\left[
\stac{(1/2)}{K_{\theta}^{\;\mu}}-\frac{1}{\e M^4}
\left(\partial_{\theta} \Sigma^{(0)}-\e A_{\theta}^{(0)}\right)n_y\stac{(1/2)}{F^{y\mu}}
\right]\right|_{y=y_{\pm}\mp\epsilon}=0. \label{1/2bc}
\end{eqnarray}
Thus, it is now clear that both
$C_1^{\nu}$ and $C_2^{\nu}$ vanish in the bulk.

To sum up, we have shown in this appendix that
\begin{eqnarray}
K_{\theta}^{\;\mu}&=&\varepsilon^{3/2}\stac{(3/2)}{K_{\theta}^{\;\mu}}+\cdots,
\\
F^{y\mu}&=&\varepsilon^{3/2}\stac{(3/2)}{F^{y\mu}}+\cdots,
\end{eqnarray}
and that
\begin{eqnarray}
\left(\partial^{\mu}\Sigma-\e A^{\mu}\right)^{(1/2)}=0
\quad\text{on the branes}.
\end{eqnarray}

\section{Solving the first order evolution equations and Hamiltonian constraint}\label{app:solve}

For completeness
we present here how we can solve the evolution equations~(\ref{tp1st}),~(\ref{thth1st}),
and the Hamiltonian constraint~(\ref{ham1st}) analytically.
Eliminating $\cF$, we obtain
the coupled evolution equations for
$\stac{(1)}{K_{\lambda}^{\;\lambda}}$ and $\stac{(1)}{K_{\theta}^{\;\theta}}$:
\begin{eqnarray}
\py \!\stac{(1)}{K_{\lambda}^{\;\lambda}}+\left(\frac{2}{y}-\frac{\py f}{2f}\right)
\stac{(1)}{K_{\lambda}^{\;\lambda}}-\frac{4}{y}\stac{(1)}{K_{\theta}^{\;\theta}}
&=&0,
\label{kev1}
\\
\py\! \stac{(1)}{K_{\theta}^{\;\theta}}+
\left(\frac{10}{y}+\frac{\py f}{f}\right)\stac{(1)}{K_{\theta}^{\;\theta}}
+\left(\frac{9}{2y}+\frac{5\py f}{4f}\right)\stac{(1)}{K_{\lambda}^{\;\lambda}}
&=&\frac{3}{4y^2}\frac{L_I}{\sqrt{f}}R.
\label{kev2}
\end{eqnarray}
Then, we substitute Eq.~(\ref{kev1}) into Eq.~(\ref{kev2}) to obtain 
the single second order differential equation for $\stac{(1)}{K_{\lambda}^{\;\lambda}}$.
In terms of
\begin{eqnarray}
\Omega:=\frac{\stac{(1)}{K_{\lambda}^{\;\lambda}}}{\sqrt{f}\,\py\!\left(\sqrt{f}/y\right)},
\label{def:omega}
\end{eqnarray}
the equation can be written in a simple form:
\begin{eqnarray}
\py^2\Omega-\frac{\py^2\cI}{\py\cI}\py\Omega=\frac{3}{y^3 f\,\py\!\left(\sqrt{f}/y\right)}L_IR,
\label{eq:omega}
\end{eqnarray}
where
\begin{eqnarray}
\py\cI(y):=\frac{1}{y^{16}(\sqrt{f}/y)^3[\py (\sqrt{f}/y)]^2}.
\end{eqnarray}
The general solution to Eq.~(\ref{eq:omega}) is given by
\begin{eqnarray}
\Omega=
C_1(x)+C_2(x)\cI(y)
+3L_IR\int^y\! d\bar y\;\partial_{\bar y}\cI(\bar y)\int^{\bar y}\!
\frac{d\tilde y}{\tilde y^3 f\partial_{\tilde y} (\sqrt{f}/\tilde y )  \partial_{\tilde y}\cI  (\tilde y) }.
\end{eqnarray}
Eqs.~(\ref{kev1}) and~(\ref{def:omega}) allow one to write
the extrinsic curvature in terms of $\Omega$.
Then $\cF$ can also be written in terms of $\Omega$ using the Hamiltonian constraint.

%===================================================%

%---------   References   ---------%

%---------   References   ---------%

\end{document}